\documentclass[fleqn,12pt]{article}

\usepackage{amoretti_bormio2005}
\usepackage{graphicx}

\def\Hbar{$\bar{H}\ $}
\def\pbar{$\bar{p}\ $}
\def\ebar{$e^+\ $}

\title{\Large\bf Cold-Antimatter Physics\footnote{Talk presented at the XLIII International Meeting on Nuclear Physics, Bormio (Italy), March 13-20 (2005). Corresponding author (marco.amoretti@ge.infn.it).}}

\author{
\underline{M.~Amoretti}\address[INFNGE]{Istituto Nazionale di Fisica Nucleare, 
Sezione di Genova, 16146 Genova, Italy},
C.~Amsler\address[ZURICH]{Physik-Institut, Z\"{u}rich University, 8057 Z\"{u}rich, Switzerland}, 
G.~Bonomi\address[CERN]{Physics Department, CERN, 1211 Geneva 23, Switzerland}, 
P.~D.~Bowe\address[AAHRUS]{Department of Physics and Astronomy, University of Aarhus, 8000 Aarhus C, Denmark},
C.~Canali\addressmark[INFNGE]$^,$\address[UNIGE]{Dipartimento di Fisica, Universit\`{a} di Genova, 16146 Genova, Italy}, 
C.~Carraro\addressmark[INFNGE]$^,$\addressmark[UNIGE], 
C.~L.~Cesar\address[BRASIL]{Instituto de Fisica, Universidade Federal do Rio de Janeiro, Rio de Janeiro 21945-970, Brazil}, 
M.~Charlton\address[SWANSEA]{Department of Physics, University of Wales Swansea, Swansea SA2 8PP, UK}, 
M.~Doser\addressmark[CERN],
A.~Fontana\address[INFNPV]{Istituto Nazionale di Fisica Nucleare Sezione di Pavia, 27100 Pavia, Italy}$^,$\address[UNIPV]{Dipartimento di Fisica Nucleare e Teorica, Universit\`{a} di Pavia, 27100 Pavia, Italy}, 
M.~C.~Fujiwara\address[JAPAN]{Department of Physics, University of Tokyo, Tokyo 113-0033, Japan}$^,$\address[RIKEN]{Atomic Physics Laboratory, RIKEN, Saitama 351-0198, Japan},
R.~Funakoshi\addressmark[JAPAN], 
P.~Genova\addressmark[INFNPV]$^,$\addressmark[UNIPV], 
J.~S.~Hangst\addressmark[AAHRUS],  
R.~S.~Hayano\addressmark[JAPAN], 
I.~Johnson\addressmark[ZURICH], 
L.~V.~J{\o }rgensen\addressmark[SWANSEA], 
A.~Kellerbauer\addressmark[CERN], 
V.~Lagomarsino\addressmark[INFNGE]$^,$\addressmark[UNIGE],
R.~Landua\addressmark[CERN], 
E.~Lodi~Rizzini\addressmark[INFNPV]$^,$\address[BRESCIA]{Dipartimento di Chimica e Fisica per l'Ingegneria e per i Materiali, Universit\`{a} di Brescia, 25123 Brescia, Italy}, 
M.~Macr\'\i\addressmark[INFNGE], 
N. Madsen\addressmark[AAHRUS], 
G. Manuzio\addressmark[INFNGE]$^,$\addressmark[UNIGE], 
D.~Mitchard\addressmark[SWANSEA], 
P. Montagna\addressmark[INFNPV]$^,$\addressmark[UNIPV], 
H.~Pruys\addressmark[ZURICH], 
C.~Regenfus\addressmark[ZURICH], 
A.~Rotondi\addressmark[INFNPV]$^,$\addressmark[UNIPV], 
G.~Testera\addressmark[INFNGE], 
A.~Variola\addressmark[INFNGE], 
L.~Venturelli\addressmark[INFNPV]$^,$\addressmark[BRESCIA],
D.~P.~van~der~Werf\addressmark[SWANSEA], 
Y.~Yamazaki\addressmark[RIKEN], and
N.~Zurlo\addressmark[INFNPV]$^,$\addressmark[BRESCIA]
\\ (ATHENA Collaboration)
}
       
\begin{document}

\maketitle

\begin{quote}
\centerline{\bf Abstract} 
The CPT theorem and the Weak Equivalence Principle are foundational principles on which the standard description of the fundamental interactions is based. The validity of such basic principles should be tested using the largest possible sample of physical systems. Cold neutral antimatter (low-energy antihydrogen atoms) could be a tool for testing the CPT symmetry with high precision and for a direct measurement of the gravitational acceleration of antimatter. After several years of experimental efforts, the production of low-energy antihydrogen through the recombination of antiprotons and positrons is a well-established experimental reality. An overview of the ATHENA experiment at CERN will be given and the main experimental results on antihydrogen formation will be reviewed.
\end{quote}

\section{Introduction}
\label{sec:intro}

Two experiments, ATHENA~\cite{athenaNIMapparatus} and ATRAP~\cite{Gabrielse1999}, have been set up at the CERN Antiproton Decelerator (AD)~\cite{AD} since 1998 with the goal of producing sufficient amounts of antihydrogen ($\bar{H}$) to ultimately allow high precision comparison of the interaction of hydrogen ($H$) and antihydrogen with electromagnetic and gravitational fields. The production of large amounts of \Hbar was first demonstrated by ATHENA~\cite{athenaNATURE} and later by ATRAP~\cite{Gabrielse2002}, using similar schemes for antihydrogen production but different detection techniques. 

\section{Motivations}
\label{sec:motivations}

\subsection{CPT}
\label{sec:cpt}
Testing fundamental symmetries is an important objective in physics. Invariance of physical laws under the combined operations, taken in any order, of charge conjugation, parity, and time reversal (CPT), is guaranteed in local quantum theories of point-like particles in  space time by the CPT theorem~\cite{Pauli1957,Luders1957} under assumptions including Lorentz invariance and unitarity. These assumptions, however, are not implicit in some classes of theories beyond the Standard Model. Recently, there is growing interest in CPT and Lorentz violations, and this is due in part to the development by Kosteleck\'{y} and co-workers of a phenomenological extension of the Standard Model (SME)~\cite{Kostelecky} that incorporates the possibility of such violations. Possible CPT violations involving quantum gravity were discussed by Wald~\cite{Wald1980} and, more recently, Ellis and co-workers have proposed scenarios of violation in this context~\cite{Ellis1996}.

Since CPT transforms an elementary particle into its antiparticle, their fundamental properties such as mass, charge, and magnetic moment, are either exactly equal or exactly opposed. This predestines antimatter for tests of CPT symmetry.  
There exist numerous experimental tests of CPT invariance~\cite{PDG2004}, of which the most often quoted is that of the neutral kaon relative mass difference at the level of $10^{-18}$. Note however that some authors~\cite{Kobayashi1992,Bigi2001} have questioned the significance of dividing the possible mass difference with the mass itself. Moreover, some figures of merit used to measure possible CPT violation could be inappropriate. For example in the framework of Kosteleck\'{y}'s SME, the relative difference between the anomalous magnetic moment of the electron and the positron ($e^+$) is zero even though CPT is broken~\cite{Kostelecky1997}.

Given the fundamental importance of CPT symmetry, it should be tested in all particle sectors where precision results can be obtained. Due to the fact that atomic spectroscopy on the transition between the ground and first excited states (1S–-2S) of hydrogen has been carried out to $10^{-14}$ relative precision~\cite{Niering2000}, this transition is also being targeted for CPT tests with hydrogen and antihydrogen. In the SME framework,  1S-–2S spectroscopic signals for Lorentz and CPT violation appear for transitions involving spin-mixed states in $H$ or \Hbar atoms confined in a magnetic trap~\cite{Kostelecky1999}.

\subsection{WEP}
\label{sec:wep}

In addition to atomic spectroscopy, tests of the validity of the Weak Equivalence Principle (WEP) for antimatter are being considered. 

In the gravitational sector, no direct measurements exist for the gravitational force on antimatter. Therefore even a measurement performed with relatively low precision ($\sim 0.1$\%) will represent an important milestone. The indirect limits~\cite{Fischbach1999} that can be obtained on the validity of the WEP for antimatter using data provided by experiments on matter, together with general physics principles, set the scale for the ultimate experimental precision that has to be reached.  Arguments related to the effect of virtual electron-positron pairs in atoms with different nuclear charges, together with the actual experimental limits on the independence of the gravitational force on the body compositions, exclude any difference in the gravitational acceleration $g$ of matter and antimatter larger than about $10^{-6}$--$10^{-7}$. Several critiques of this argument have been discussed in the literature, emphasizing the need of a direct measurement~\cite{Fischbach1999,NietoGoldman}.

Experimental WEP tests with antihydrogen could be carried out by gravity interferometry.
Phillips~\cite{Philips1997} has proposed to use a Mach-Zehnder interferometer to measure the displacement of a horizontal beam of antihydrogen atoms under the influence of gravity. 
Relative accuracies $\Delta g/g$ of a few percent or better should be possible.
Alternative schemes could be developed by analogy with the interferometric technique used to study gravity effects on ordinary matter with atomic fountains. Direct measurements of the gravitational acceleration of cold atoms in the Earth's gravitational field using these techniques have been shown to achieve a sensitivity of $10^{-9}$~\cite{Peters1999}. Once a sample of very cold (sub-meV) antiatoms is available, the same precision could be obtained for the gravitational interaction of antimatter.

\section{Experimental Apparatus}
\label{sec:apparatus}
\begin{figure}
\centering
\resizebox{0.95\textwidth}{!}{\includegraphics{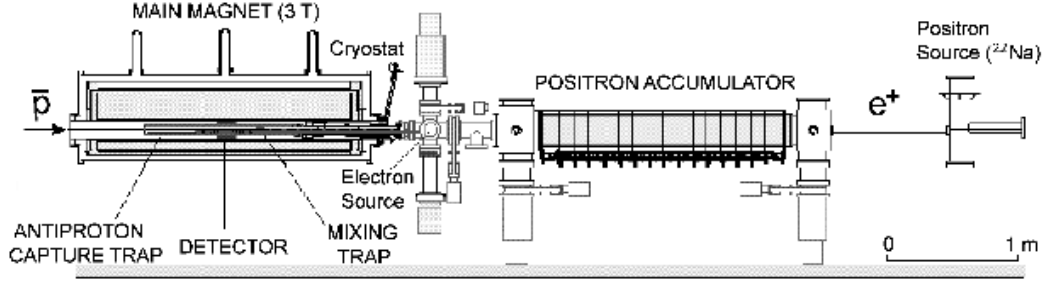}}
\caption{\label{fig:1} Scheme of the ATHENA apparatus for the production and detection of antihydrogen.}
\end{figure}

The ATHENA apparatus~\cite{athenaNIMapparatus}  consists of three main electromagnetic traps for charged particles: the trap used to catch and cool the antiprotons ($\bar{p}$), the  positron accumulator, and the trap for mixing the particles located in the middle between the two previous traps (see Fig.~\ref{fig:1}). The antiproton and mixing traps are placed inside an ultra-high vacuum cryostat immersed in a 3 T superconducting magnetic field. The positron accumulation trap is located inside a room temperature vacuum chamber in a 0.14 T magnetic field. An imaging particle detector~\cite{Regenfus2003}, used for identification of the \Hbar annihilation products, surrounds the mixing trap  [Fig.~\ref{fig:2}(a)].

The antiproton capture trap is a multi-electrode cylindrical Penning-Malmberg trap in which antiprotons supplied by the AD are trapped in a high-voltage potential well ($\sim 5$ kV), after being slowed down by means of a degrader foil. These trapped antiprotons are cooled by collisional interaction with preloaded electrons~\cite{nnp2002}, which in turn self-cool via the emission of synchrotron radiation. In a standard mixing cycle, antiprotons from two or three AD spills are accumulated in the catching trap before they are transferred to the adjacent mixing trap. About 1--$2 \cdot 10^4$ antiprotons (per cycle) were transfered and used for mixing with positrons. 

Positrons are obtained from a radioactive source ($^{22}$Na) and are moderated by a frozen neon film. Their trapping and accumulation are achieved with the help of nitrogen buffer gas~\cite{Murphy1992,Jorgensen2002}, which provides the dissipative process necessary for trapping the continuous flow of positrons. In the positron accumulator about $1.5\cdot 10^8$ positrons are accumulated in cycles of roughly 5 minutes. They are then transfered into the mixing trap with an overall efficiency of about 50\%; here they reach a thermal equilibrium with the surrounding environment by emission of synchrotron radiation in the 3 T field. The result is spheroidal positron plasma with a density of about ($10^8$--$10^9$ cm$^{-3}$), a maximum length of about 3.5 cm and a typical radius of $0.1$--$0.2$ cm~\cite{modes}.

\section{First Antihydrogen Production and Detection}
\label{sec:hbar}
\begin{figure}
\centering
\resizebox{0.95\textwidth}{!}{\includegraphics{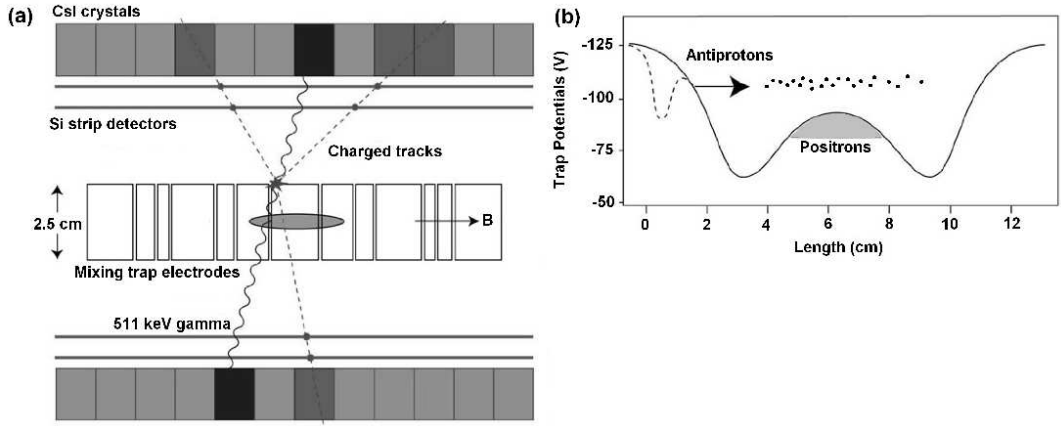}}
\caption{\label{fig:2} (a) Schematic diagram, in axial section, of the ATHENA mixing trap and \Hbar detector with a typical annihilation event. (b) The trapping potential on the axis of the mixing trap is plotted against length along the trap. The dashed line is the potential immediately before antiproton injection while the solid line is the potential during mixing.}
\end{figure}

The technique used to mix the \pbar and the \ebar clouds is based on the so-called ``nested'' potential configuration~\cite{Gabrielse1988}, which permits simultaneous confinement of oppositely charged particles [Fig.~\ref{fig:2}(b)]. After the transfer of the two species into the mixing region, the antiprotons are injected into the positron plasma initiating the interaction process. An antiproton rapidly loses its energy via Coulomb collisions inside the positron plasma and eventually captures a positron producing an \Hbar atom.

\begin{figure}
\centering
\resizebox{0.95\textwidth}{!}{\includegraphics{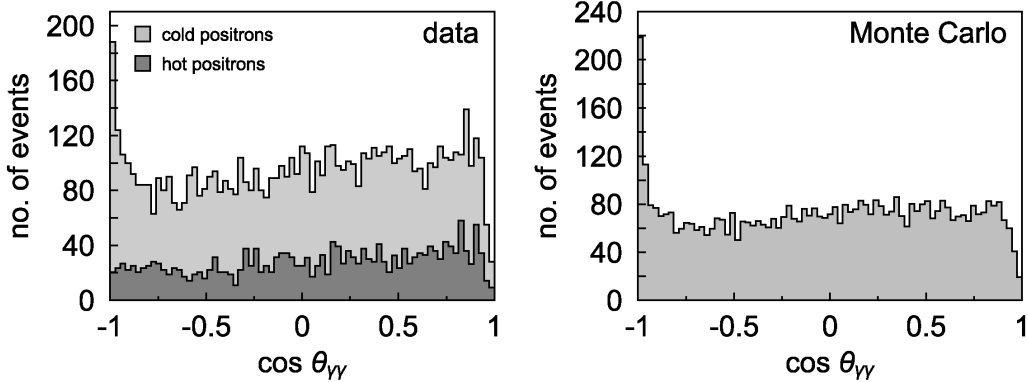}}
\caption{\label{fig:3} \small Distribution of the opening angle between two 511 keV gamma rays, seen from the reconstructed vertices of antiproton annihilations in the case of antiproton mixing with cold and hot positrons.}
\end{figure}

The formed atoms are neutral and are not confined by the electromagnetic fields of the charged-particle trap. They drift toward the walls and annihilate in interaction with matter [see Fig.~\ref{fig:2}(a)]. Antihydrogen atoms are identified by detecting simultaneous annihilation (within $\sim 5 \ \mu$s) of antiprotons and positrons at the same place (vertex reconstruction $\sigma = \pm 4$ mm). Antiprotons annihilate into several charged or neutral particles (mostly pions), and the annihilation vertices are reconstructed by tracking the charged trajectories with two layers of double-sided silicon microstrip detectors. For each vertex, we search for 'clean' evidence of 511-keV photons in the crystal data. A charged-particle hit in a crystal, or an outer-layer silicon hit lying in the footprint of a crystal, excludes that particular crystal and its eight nearest neighbors. Next, we demand that exactly two of the remaining crystals have hits in an energy window around 511 keV, and that there are no hits of any energy adjacent to these two crystals. Energy calibration data, measured for each individual crystal, are used in this test. To search for antihydrogen in the sample of events having a vertex and two 'clean' photons, we consider the opening angle $\theta_{\gamma\gamma}$ between the lines connecting the vertex point to the geometric centers of the two hit crystals. For an \Hbar event, this angle should be $180^\circ$ (or $\cos \theta_{\gamma\gamma} = -1$). The opening angle distribution for reconstructed events detected during standard mixing cycles (also called ``cold mixing'' cycles) is shown in Fig.~\ref{fig:3}(left)~\cite{athenaNATURE}. For the real \Hbar events, there should be a peak at $\cos \theta_{\gamma\gamma} \simeq -1$, and indeed this is what we experimentally observe.

The background was carefully studied in several ways. Measurements without positrons and only with  antiprotons annihilating on the electrode wall were taken. The storage time of the antiprotons without the interaction with the positrons can reach several hours~\cite{Fujiwara2001} and we induce antiproton radial loss using different procedures, e.g., by injecting a low-energy (few tens of eV) electron beam throught the antiproton cloud. In addition standard mixing data were analyzed with the photon energy cut displaced. In both the background cases, no peak at $\cos(\theta_{\gamma\gamma}) \simeq -1$ is observed, as expected~\cite{athenaNATURE}. The three-dimensional imaging capability of the antiproton annihilation~\cite{athenaIMAGING}, as well as high angular resolution for photon detection with segmented crystals, were essential in discriminating against the angular-uncorrelated photon background, which comes predominantly from the decay and the subsequent electromagnetic shower of neutral pions. 

An additional measurement of the background and a confirmation of the \Hbar production was obtained by mixing \pbar with a cloud of ``hot'' positrons. The temperature of the positron plasma was increased by applying a radio-frequency signal~\cite{modes}. 
When the positron plasma is heated to several 1000 K, the \Hbar production is suppressed and only a much lower number of events, from antiproton annihilations with trapped residual gas ions or neutral contaminants, is
detected. As shown in Fig.~\ref{fig:3}(left), the peak for an opening angle of $180^\circ$ disappears when the positrons are heated. Figure~\ref{fig:3} also illustrates the excellent agreement of the cold mixing data with a Monte Carlo simulation if a suitable background is subtracted (see the following discussion).

\section{Recent Results}
\label{sec:results}
For precise antimatter studies, it is not sufficient to merely produce large numbers of antihydrogen atoms. Knowledge of the temperature and kinetic-energy distributions of the \Hbar is required in order to estimate the fraction of antiatoms which may be trapped. The \Hbar atoms must also be produced in a well-defined internal quantum state, preferebly the ground state. Since the first production of \Hbar atoms, the main challenges have been to investigate the parameters that govern efficient \Hbar formation and its internal and external properties.

\subsection{Antihydrogen Production 2002/2003}
\label{subsec:results}
As a prerequisite for any quantitative studies on \Hbar formation, the data analysis must allow a precise determination of the number of \Hbar atoms produced~\cite{athenaHIGHRATE}. 

In order to achieve this, one or several observables, such as the radial vertex distribution or the $2\gamma$ opening angle distribution, can be considered as a linear combination of a pure \Hbar signal (Monte-Carlo simulation of annihilations on the trap electrodes) and background. Since the background is expected to be mainly due to \pbar annihilations with residual gas, it can be represented by the signal obtained from hot-mixing runs.
The total \Hbar production of 2002 and 2003 obtained in this way is summarized in Tab.~\ref{tab:1}. It shows that ATHENA produced more than $10^6$ \Hbar atoms in 2002 and 2003 combined and that the production efficiency in terms of captured antiprotons from the AD is between 10 and 20\%.

A complete analysis of the data, together with more detailed Monte Carlo simulations, also showed that the instantaneous trigger rate from the silicon detector is a good proxy for \Hbar production, with about 70\% of all triggers over the entire mixing cycle due to annihilating \Hbar atoms~\cite{athenaHIGHRATE}.

\begin{table}
\centering
\begin{tabular}{lcc}
		  \hline \hline
			                               &	Cold mixing 2002 & Cold mixing 2003 \\
			\hline
		  Total no. of cycles            & 341               & 416 \\
			Cycle duration 			           & 180 s             & 70 s \\
			Total mixing time              & 17.1 h            & 8.1 h \\
			Injected \pbar $^\dag$         & $2.92 \cdot 10^6$ & $5.07 \cdot 10^6$ \\
			Produced \Hbar $^\dag$         & $4.94 \cdot 10^5$ & $7.04 \cdot 10^5$ \\
			Production efficiency$\ ^\dag$ & 16.9\%            & 13.9\% \\
			Avg. \Hbar production rate     & 8.0(4) Hz         & 24.2(1.3) Hz \\
			\Hbar fraction of signal       & 65(5)\%           & 74(3)\% \\
		  \hline \hline
			\end{tabular}
\caption{\label{tab:1} \small Comparative summary of ATHENA antihydrogen production in 2002 and 2003. (Note: $^\dag$5\% relative uncertainty)}
\end{table}

\subsection{\Hbar Production Temperature Dependence}
\label{sec:temp}
The formation of \Hbar by direct capture of a positron into an atomic orbit around an antiproton does not simultaneously conserve energy and momentum. The involvement of a third particle is needed in order to respect these conservation laws. That particle can either be a photon in the case of (spontaneous) radiative recombination (SRR)~\cite{SRR} or a second positron in three-body recombination (TBR)~\cite{TBR}. These two processes are predicted to have vastly different cross-sections and recombination rates, with TBR expected to be the dominant process for ATHENA's experimental conditions. The most important difference with a view to precision studies lies in the fact that SRR populates low-lying states ($n < 10$) and TBR highly excited Rydberg states ($n >> 10$). On the hypothesis of thermal equilibrium between positrons and antiprotons, the two mechanisms also exhibit different dependencies on the positron temperature (SRR: $\propto T^{-0.63}$; TBR: $\propto T^{-4.5}$),
which can allow them to be distinguished.

\begin{figure}
\centering
\resizebox{0.95\textwidth}{!}{\includegraphics{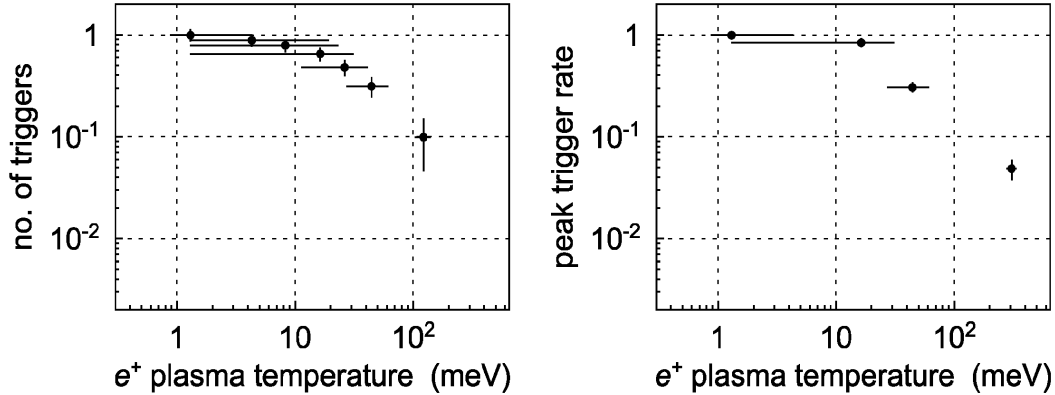}}
\caption{\label{fig:4} \small Dependence of the background-corrected integrated total number of charged-particle triggers per mixing cycle (left) and the peak trigger rate (right) on the positron plasma temperature. The number of triggers and trigger rate have been normalized to cold mixing values (e$^+$ temperature $\simeq 15$ K). Note the logarithmic scale.}
\end{figure}

In order to determine the temperature dependence of \Hbar production, we have performed mixing cycles with RF heating at various amplitudes applied to the positron plasma~\cite{athenaTEMPERATURE}. The positron temperature increase was measured using a diagnostics system~\cite{modes} from the resonant excitation and detection of the axial positron plasma modes. In Fig.~\ref{fig:4}, the background-corrected integrated number of triggers (left) and peak trigger rate (right) as possible proxies for \Hbar production are shown as a function of the positron temperature, assuming an equilibrium temperature for cold mixing of 15 K. Neither of these plots shows the characteristics of a simple power law (a straight line in these logarithmic plots), but a best fit to the data yields a behavior of the form $\propto T^{-0.7\pm 0.2}$, close to that expected from radiative recombination. However, the observed event rates are between 1 and 2 orders of magnitude higher than expected for this recombination process.

\subsection{Antihydrogen Spatial Distribution}
\label{sec:spatial}

Using the \Hbar annihilation detector, it is also possible to investigate the spatial distribution of the emerging \Hbar atoms~\cite{athenaSPATIAL}. The main result of this analysis is that distribution is independent of the \ebar temperature and enhanced in the axial direction. The latter assumes homogeneous formation throughout the positron plasma, and rotation of the antiproton with the positrons. We argue that this indicates that \Hbar is not formed under conditions of thermal equilibrium between positrons and antiprotons. This observation could also explain why the production does not show a simple power law dependence on the \ebar temperature. The lower limit of the \Hbar formation temperature (or equivalentely, kinetic energy) is $\sim 150$ K in the direction parallel to the main magnetic field and $\sim 15$ K perpendicular to it. 

If, indeed, the \Hbar are formed before equilibrium as our results indicate, it could be an obstacle to the trapping of $\bar{H}$, and thus precision spectroscopy. Typical neutral traps have depths of about 1 K. If \Hbar atoms are formed at some hundreds of K, very few will be trapped. Lowering the positron temperature may not help as formation is fast compared to the cooling~\cite{athenaCOOLING}. To produce cold \Hbar, it is thus necessary to have cold \pbar before mixing with $e^+$. A possible solution could be to invert the current mechanism for formation. That is, trap the \pbar in the center of the nested trap, possibly with some electrons to keep them cold, and then pass \ebar through them, reinjecting them regularly as has already been done with \pbar~\cite{GabrielseRF}. Positronium formation or axial separation of \pbar and \ebar are both potential obstacles to this scheme. An alternative scheme that should also produce \Hbar at or close to ambient temperature would be to form \Hbar by $\bar{p}$--positronium collisions~\cite{Humberston1987,Charlton1990}.

\subsection{Laser Stimulated Recombination}
\label{sec:laser}

At the end of the 2003 run the apparatus was modified to allow the introduction of laser light into the system. During the 2004 run, the main experimental effort has been the attempt to stimulate radiative recombination with the use of a CO$_2$ laser. The laser source was tunable in a wavelength range between $9.5$ $\mu$m and  11.2 $\mu$m. The maximum laser beam power was 30 W. The beam waist in the mixing region was about 2 mm with a typical peak intensity of 160 W cm$^{-2}$ at 10 W power.

The wavelength with the expected maximum enhancement in \Hbar production is 10.96 $\mu$m, corresponding to the transition from the continuum to the $n=11$ quantum state. The expected stimulated formation rate was about 60 Hz under equilibrium conditions at 15 K. The analysis of the collected data is still in progress; however, no obvious enhancement in \Hbar production was observed in the experiment.

\section{Conclusions}
\label{sec:concl}
With the first production of copious amounts of cold antihydrogen, many of the challenges on the way to high-precision CPT tests with antimatter have been surmounted, but many still remain. Future high-precision spectroscopic and interferometric measurements on antimatter atoms are contingent upon the ability to confine neutral \Hbar atoms and possibly to cool them with Lyman-$\alpha$ lasers. Our results on the temperature dependence of \Hbar production suggest on the one hand that an appreciable fraction of the antihydrogen may be produced in low-lying states accessible to precision atomic spectroscopy. On the other hand, recombination possibly sets in before complete thermalization of the antiprotons, thereby reducing the fraction of produced antihydrogen that can be confined in a magnetic trap. Further studies on \Hbar production in a nested Penning trap are required to clarify these points. In parallel, tests with ordinary matter on the simultaneous confinement of charged and neutral particles in electro-magnetic traps are being carried out in order to establish parameters for the efficient preparation of trapped antihydrogen for symmetry tests.

\section{Acknowledgments}
\label{sec:acknow}
The authors acknowledge J. Rochet, S. Bricola, G. Sobrero, and P. Chiggiato for their valuable technical support, and CERN's PS and AD crew for providing the excellent antiproton beam.  This work was supported by INFN (Italy), FAPERJ (Brasil), MEXT (Japan), SNF (Switzerland), NSRC (Demark), EPSRC (UK), and the European Commission.


{\small
\begin{thebibliography}{99}

\bibitem{athenaNIMapparatus}
ATHENA Collaboration (M. Amoretti {\em et al.}), Nucl. Instrum. Methods A {\bf 518}, 679 (2004).

\bibitem{Gabrielse1999}
G. Gabrielse {\em et al.}, Phys. Lett. B 455, 311 (1999).

\bibitem{AD}
S. Maury, Hyp. Int. {\bf 109}, 43 (1997).

\bibitem{athenaNATURE}
ATHENA Collaboration (M. Amoretti {\em et al.}), Nature {\bf 419}, 456 (2002).
 
\bibitem{Gabrielse2002}
G. Gabrielse {\em et al.}, Phys. Rev. Lett. {\bf 89}, 213401 (2002).

\bibitem{Pauli1957}
W. Pauli, Il Nuovo Cimento {\bf 6}, 6 (1957).

\bibitem{Luders1957}
G. Luders, Ann. Phys. {\bf 2}, 1(1957) .

\bibitem{Kostelecky}
See, e.g., D. Colladay and V.~A. Kosteleck\'{y}, Phys. Rev. D {\bf 55}, 6760 (1997). A complete list of references can be found at
{\tt http://www.physics.indiana.edu/$\sim$kostelec}.

\bibitem{Wald1980}
R.~M. Wald, Phys. Rev. D {\bf 21}, 2742 (1980).

\bibitem{Ellis1996}
J. Ellis, J. Lopez, N.~E. Mavromatos, D.~V. Nanopoulos, Phys. Rev. D {\bf 53}, 3846 (1996).

\bibitem{PDG2004}
S. Eidelman {\em et al.} (Particle Data Group), Phys. Lett. B {\bf 592}, 1 (2004).

\bibitem{Kobayashi1992}
M. Kobayashi, A.~I. Sanda, Phys. Rev. Lett. {\bf 69}, 3139 (1992).

\bibitem{Bigi2001}
I.~I. Bigi, Nucl. Phys. A {\bf 692}, 227c (2001).

\bibitem{Kostelecky1997}
R. Bluhm, V.~A. Kosteleck\'{y}, and N. Russell, Phys. Rev. Lett. {\bf 79}, 1432 (1997).

\bibitem{Niering2000}
M. Niering {\em et al.}, Phys. Rev. Lett. {\bf 84}, 5496 (2000).

\bibitem{Kostelecky1999}
R. Bluhm, V.~A. Kosteleck\'{y}, and N. Russell, Phys. Rev. Lett. {\bf 82}, 2254 (1999).

\bibitem{Fischbach1999}
E. Fischbach and C.~L. Talmadge, {\it ``The search for Non-Newtonian Gravity''}, Springer Verlag (New York, 1999).

\bibitem{NietoGoldman}
M.~M. Nieto and T. Goldman, Phys. Rep. {\bf 205}, 221 (1991); Phys. Rep. {\bf 216}, 343 (1992).

\bibitem{Philips1997}
T.~J. Philips, Hyp. Int. {\bf 109}, 357 (1997).

\bibitem{Peters1999}
A. Peters, K.~Y. Chung, and S. Chu, Nature {\bf 400}, 849 (1999).

\bibitem{Regenfus2003}
C. Regenfus, Nucl. Instrum. Methods A {\bf 501}, 65 (2003). 

\bibitem{nnp2002}
M. Amoretti, C. Carraro, V. Lagomarsino, G. Manuzio, G. Testera, and A. Variola, in {\em Non-Neutral Plasma Physics IV}, edited by F. Anderegg, L. Schweikhard, and C.~F. Driscoll, p. 45 (CP606, American Institute of Physics, New York, 2002).

\bibitem{Murphy1992}
T.~J. Murphy and C.~M. Surko, Phys. Rev. A {\bf 46}, 5696 (1992).

\bibitem{Jorgensen2002}
L.~V. J{\o }rgensen {\em et~al.}, in {\em Non-Neutral Plasma Physics IV}, edited by F. Anderegg, L. Schweikhard, and C.~F. Driscoll, p. 35 (CP606, American Institute of Physics, New York, 2002).

\bibitem{modes}
ATHENA Collaboration (M. Amoretti {\em et al.}), Phys. Rev. Lett. {\bf 91}, 055001 (2003);
Phys. Plasmas {\bf 10}, 356 (2003);  in {\em Non-Neutral Plasma Physics V}, edited by M. Schauer, T. Mitchell, R. Nebel, p. 121 (CP692, American Institute of Physics, New York, 2003).

\bibitem{Gabrielse1988}
G. Gabrielse, S.~L. Rolston , L. Haarsma, and W. Kells, Phys. Lett. A {\bf 129}, 38 (1988).

\bibitem{Fujiwara2001}
M.~C. Fujiwara {\em et~al.}, Hyp. Int. {\bf 138}, 153 (2001). 

\bibitem{athenaIMAGING}
ATHENA Collaboration (M.~C. Fujiwara {\em et al.}), Phys. Rev. Lett. {\bf 92}, 065005 (2004).

\bibitem{athenaHIGHRATE}
ATHENA Collaboration (M. Amoretti {\em et al.}), Phys. Lett. B {\bf 578}, 23 (2004).

\bibitem{SRR}
J. Stevefelt {\em et al.}, Phys. Rev. A {\bf 12}, 1246 (1975).

\bibitem{TBR}
M.~E. Glinsky and T.~M. O'Neil, Phys. Fluids B {\bf 3}, 1279 (1991).

\bibitem{athenaTEMPERATURE}
ATHENA Collaboration (M. Amoretti {\em et al.}), Phys. Lett. B {\bf 583}, 59 (2004).

\bibitem{athenaSPATIAL}
ATHENA Collaboration (N. Madsen {\em et al.}), Phys. Rev. Lett. {\bf 94}, 033403 (2005).

\bibitem{athenaCOOLING}
ATHENA Collaboration (M. Amoretti {\em et al.}), Phys. Lett. B {\bf 590}, 133 (2004).

\bibitem{GabrielseRF}
G. Gabrielse {\em et al.}, Phys. Rev. Lett. {\bf 89}, 233401 (2002).

\bibitem{Humberston1987}
J.~W. Humberston et al., J. Phys. B {\bf 20}, L25 (1987).

\bibitem{Charlton1990}
M. Charlton, Phys. Lett. A {\bf 143}, 143 (1990).

\end{thebibliography}
}
\end{document}